\documentclass{cmspaper}
\usepackage{upgreek}
\pdfoutput=1
\begin{document}

\begin{titlepage}


   \title{LHC results and prospects: Beyond Standard Model}

   \begin{Authlist}
        Daniel Teyssier on behalf of the ATLAS and CMS collaborations
       \Instfoot{rwth}{III Phys. Institut A, RWTH Aachen}
  \end{Authlist}

\begin{abstract}
We present the results and prospects for searches beyond the Standard Model (SM) at the LHC by the ATLAS and CMS 
collaborations. The minimal supersymmetric extension of the SM has been investigated in various 
configurations and lower limits are set on the s-particle masses. The searches for other scenarios of physics 
beyond the SM are also presented and lower limits on the mass scale are derived in a large variety of models (new heavy gauge bosons, 
extra-dimensions, compositeness or dark matter).
The prospects for physics using 300 /fb and 3000 /fb of data at the high luminosity LHC are also shown.

\end{abstract}

\conference{Talk presented at the International Workshop on Future Linear Colliders (LCWS13), 
\\Tokyo, Japan, 11-15 November 2013.}
  
\end{titlepage}

\setcounter{page}{2}

\section{Introduction}

\indent
The LHC extended the searches for physics beyond the Standard Model (SM) in an unprecedented way. The integrated 
luminosity recorded in 2011 and 2012 reached 5 /fb and 21 /fb for both the ATLAS and CMS experiments, 
for the corresponding center-of-mass energies of 7 TeV and 8 TeV. The searches for new physics are oriented to the minimal
supersymmetric extension of SM, as well as any theoretical scenario including new heavy gauge bosons, extra-dimensions, compositeness 
and dark matter, or any other signs of new physics beyond the SM.

\section{Data taking and experiments}

\subsection{Data taking at the LHC}

\indent
The years 2011 and 2012 have been extremely successful for the LHC machine. The luminosity reached the record of 
~$7.6 \times 10^{33}$ cm$^{-2}$s$^{-1}$ at the end of the pp operations. The delivered integrated luminosity was greather than 6 /fb in 2011
 and greather than 23 /fb in 2012 for both the ATLAS and CMS detectors. The data taking efficiency was better 
than 90$\%$ and the recorded integrated luminosity was of the order of 5 /fb and 21 /fb respectively in 2011 and 2012 for both 
experiments \cite{cms_lumi}\cite{atlas_lumi}, as shown in Fig.\ref{data2012}.

\begin{2figures}{hbtp}
  \resizebox{\linewidth}{0.8\linewidth}{\includegraphics{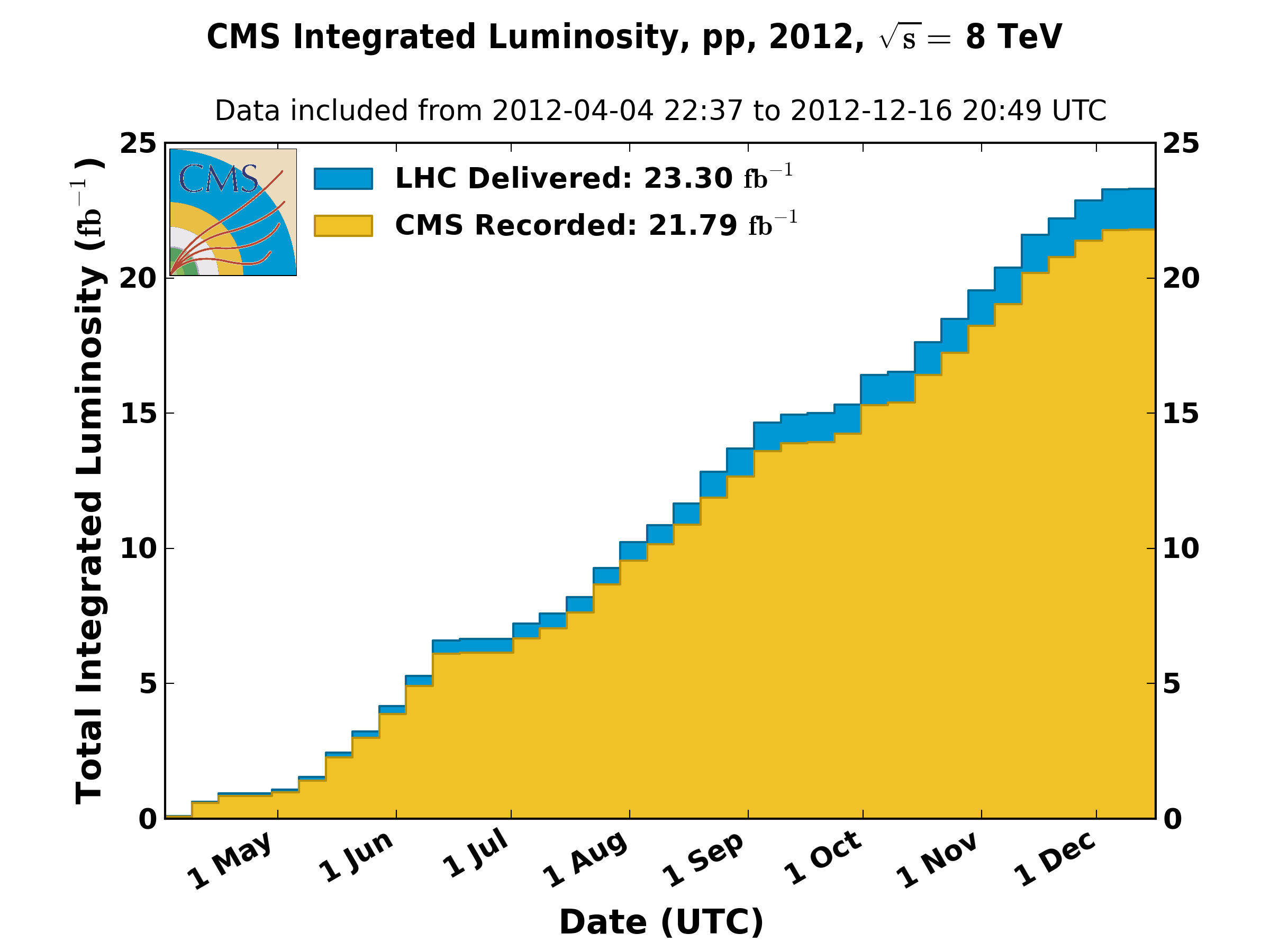}} &
  \resizebox{\linewidth}{0.8\linewidth}{\includegraphics{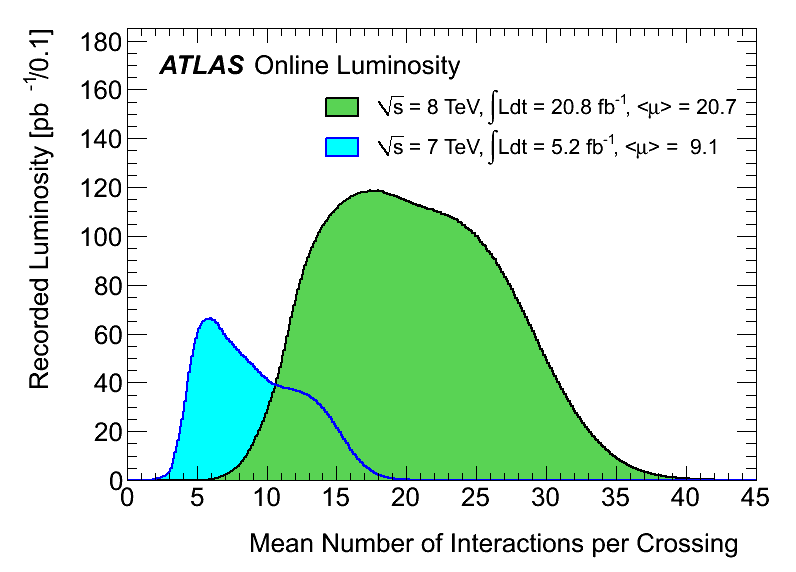}} \\
  \caption{Delivered (blue) and recorded (orange) integrated luminosity during
the 2012 data taking in CMS \cite{cms_lumi}.}
  \label{data2012} &
  \caption{Mean number of interactions per crossing in ATLAS in 2011 (blue) and 2012 (green) \cite{atlas_lumi}.}
  \label{mean_interactions} \\
\end{2figures}
 
\indent 
The energy per beam was 3.5 TeV in 2011 and was increased to 4 TeV in 2012. After the first long shutdown on-going up to the beginning of 2015,  
an intermediate phase of data taking with beams of 6.5 TeV will occur and afterwards the LHC beams will reach their nominal energy of 7 TeV.
Then the full center-of-mass energy of 14 TeV will be used to look for new physics.

\indent 
The mean number of interactions per crossing increased with the luminosity. During the 8 TeV operations, this number 
was on average 20 and even reached the value of 40 at the end of the pp operations in 2012, as shown in Fig.\ref{mean_interactions}. The pile-up is indeed a key
parameter in the physics analyses and will reach the value of 140 at the end of the High-Luminosity (HL)-LHC program. 

\subsection{ATLAS and CMS detectors}

\indent
ATLAS \cite{atlas-detector-paper} and CMS \cite{cms-detector-paper} are multi purpose detectors using a large variety of technologies 
in order to identify and trigger on electrons, photons, muons, jets and missing transverse energy 
objects \cite{atlas-tdr}\cite{cms-tdr}. Both experiments have pixel and tracker 
detectors made of semiconductors. ATLAS is using a liquid Ar electromagnetic calorimeter while CMS has chosen a PbWO4 
crystal technology. The hadronic calorimeter of ATLAS uses liquid Ar in the endcaps and tiles in the barrel. The CMS 
hadronic calorimeter is a mixture of brass and scintillating material. Both experiments are equipped with similar muon 
detectors (drift tubes, cathode strips chambers and resistive plate chambers), but the geometry of the magnetic field
(pure solenoid in CMS, internal solenoid plus toroidal in ATLAS) demands a different localization of these three 
kinds of detectors.

\section{SUSY searches}

\indent
Several arguments exist in favour of a supersymmetric (SUSY) extension of the SM.  The hierarchy problem can be solved
without too much fine-tuning and then the SUSY theory demands relatively light stop quarks 
(below 1 TeV). The implication of the discovery of the scalar boson at 126 GeV is that one stop quark can be light but the other stop should be heavier;
but this is not yet in contradiction with the previous argument. The gauge coupling unification 
at the Grand Unified Theories (GUT) scale cannot occur in the SM, but it could occur in the SUSY extension for certain set of parameters.
If R-parity is conserved ($-1^{(3(B-L)+2S)}$=+1/-1 respectively for SM/SUSY particles), a natural dark matter 
candidate is provided being the neutralino1, as the Lightest Supersymmetric Particle (LSP) will be stable. 
Finally a link to gravity is given in the SUSY theory, which is not the case in the SM.

\subsection{SUSY processes}

\indent
The different SUSY production processes can be sorted in several categories: gluino, squark and electroweakino 
(ie. chargino/neutralino/slepton) productions. Fig.\ref{susy_processes} shows the different cases and the corresponding 
expected cross sections \cite{susy_processes}. These SUSY cross sections are lower by orders of magnitude than the SM processes. The SM
background determination will be achieved using both MC simulation as well as some data driven methods.

\begin{2figures}{hbtp}
  \resizebox{\linewidth}{0.8\linewidth}{\includegraphics{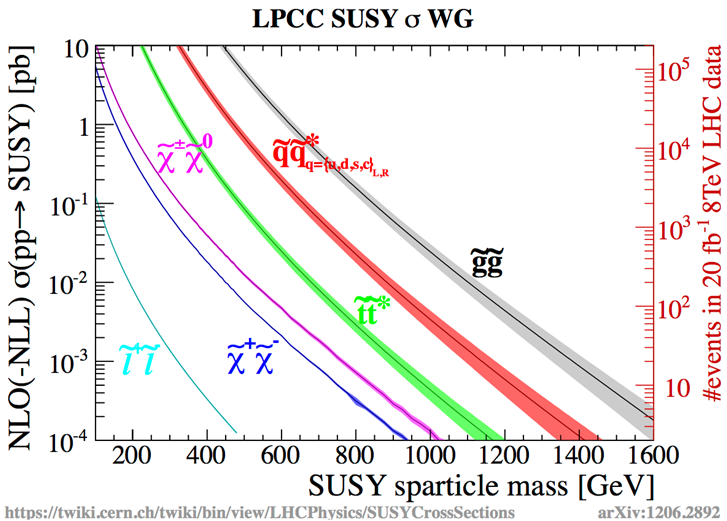}} &
  \resizebox{\linewidth}{0.8\linewidth}{\includegraphics{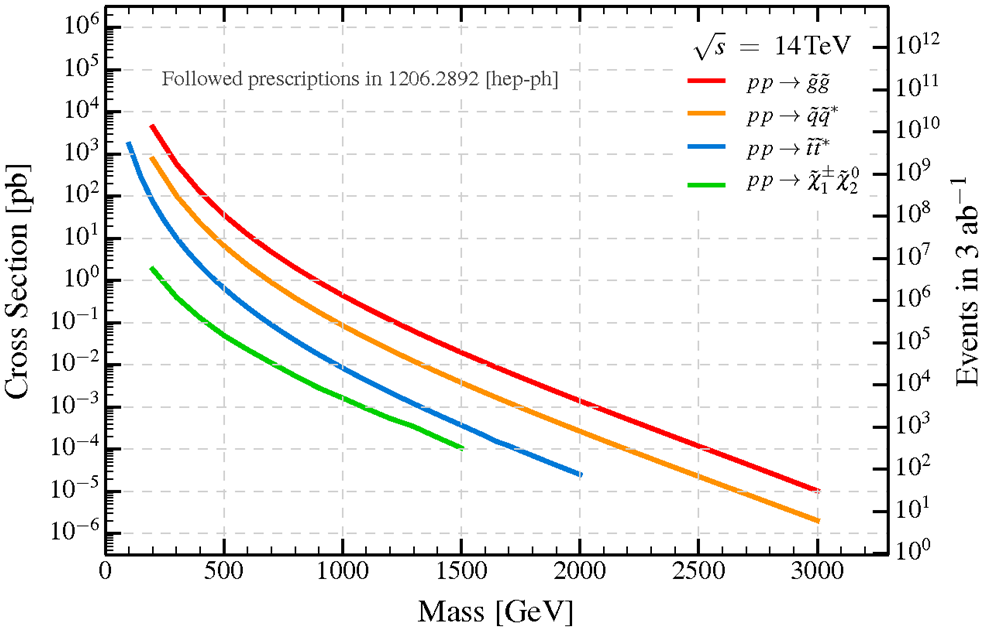}} \\
  \caption{Predicted SUSY cross sections at 8 TeV \cite{susy_processes}.}
  \label{susy_processes} &
  \caption{Predicted SUSY cross sections at 14 TeV \cite{xsec_prospects}.}
  \label{xsec_prospects} \\
\end{2figures}

\subsection{Framework for interpretation}
 
\indent
Several frameworks are commonly used to interpret the results of the SUSY searches. In order to reduce the number of
parameters in the Minimal Supersymmetric Standard Model (MSSM), which contains originally 105 parameters, some 
constraints are introduced. The mSUGRA model for instance is a 5 parameter model assuming 
the unification of sfermion masses, gaugino masses and tri-linear couplings at the GUT scale. But there are 
also other SUSY models, like GMSB (Gauge Mediated Symmetry Breaking) or pMSSM 
(phenomenological MSSM) using different constraints. Another approach is to consider only the dominant SUSY cascades and to assume the branching
ratio to 100$\%$. They are called Simplified Model Spectra (SMS). The topology is described by masses and 
cross sections. It allows to perform wider searches than the constrained models.

\subsection{Results and prospects in SMS}
 
\indent
Assuming the R-parity to be conserved, Fig.\ref{gluino_limit} shows the exclusion limit obtained using the gluino production and combining
several channels (from zero to three leptons) in the CMS experiment \cite{cms_gluino}. Fig.\ref{t1_tttt} shows an example of a schematic diagram 
of a dominant SUSY cascade (gluino production, the gluino decaying to a top pair and a neutralino), without showing the intermediate particles. 
The gluino mass had been probed up to 1.3 TeV. The limit is given at $95\%$ confidence level (CL), using all 7 TeV and 
8 TeV data. 

\begin{2figures}{hbtp}
  \resizebox{\linewidth}{0.8\linewidth}{\includegraphics{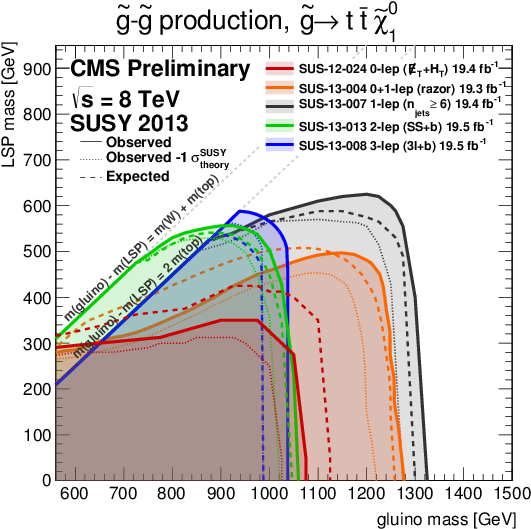}} &
  \resizebox{\linewidth}{0.5\linewidth}{\includegraphics{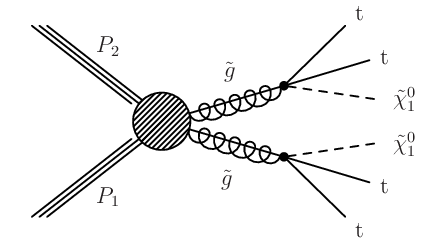}} \\
  \caption{Gluino mass limit area from CMS using all 7 TeV and 8 TeV data \cite{cms_gluino}.}
  \label{gluino_limit} &
  \caption{SUSY dominant cascade in T1 region (gluino production) in SMS model  \cite{cms_gluino}.}
  \label{t1_tttt} \\
\end{2figures}

\begin{figure}[hbtp]
 \begin{center}
  \resizebox{14cm}{!}{\includegraphics*{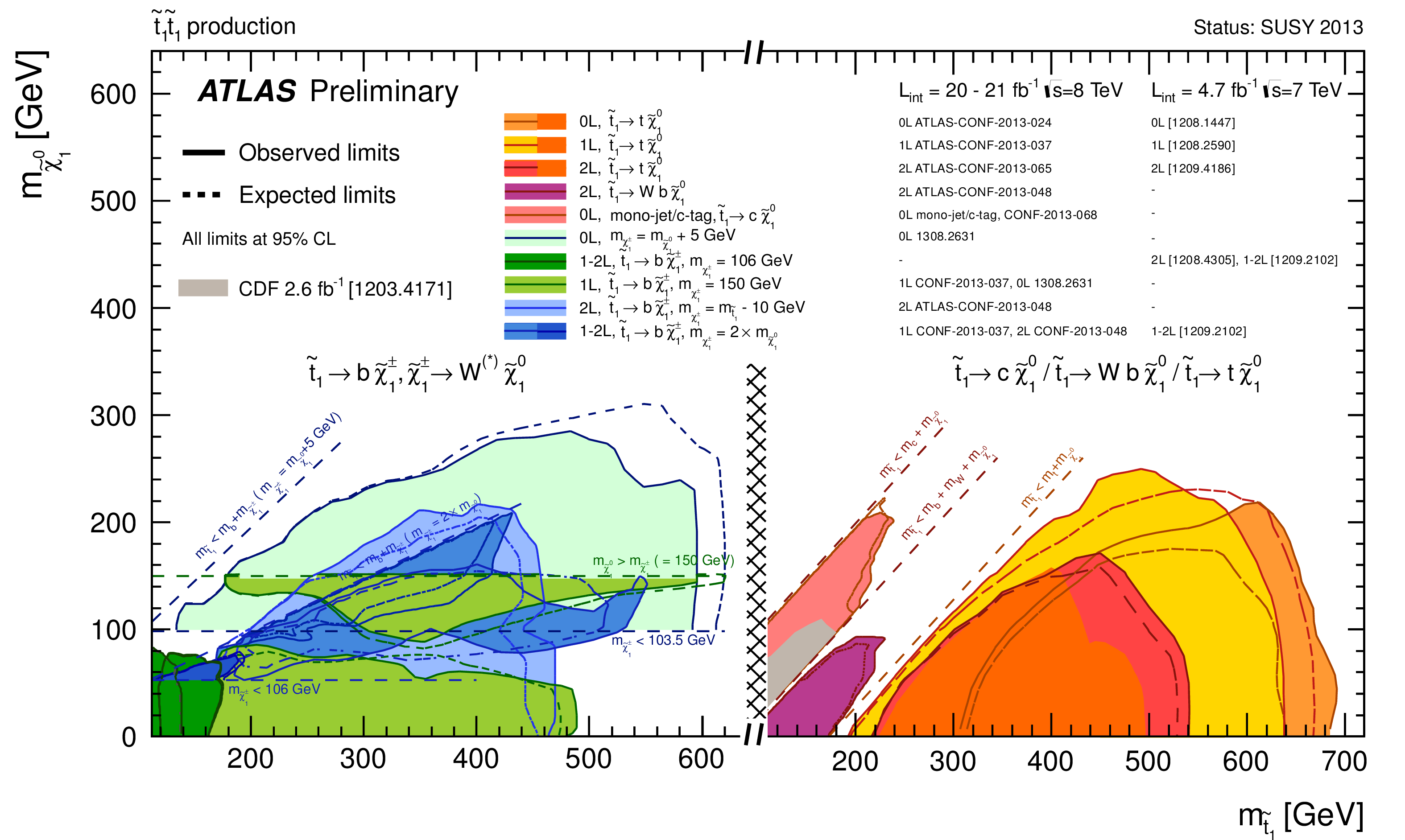}}
  \caption{Stops mass limits in ATLAS using all 7 TeV and 8 TeV data \cite{atlas_stops}.}
  \label{stops_limit}
 \end{center}
\end{figure}

\indent
Fig.\ref{stops_limit} shows the exclusion area for the stop production in ATLAS assuming the R-parity to be conserved \cite{atlas_stops}. 
Two different regions are shown depending on the final state (stops decaying to top and LSP on the left ; stops decaying to W, b quark and LSP 
on the right). The stop mass is probed 
up to roughly 700 GeV. The previous limit from the CDF experiment is also superimposed, showing the improvement by the LHC. 

\indent
The electroweakino production (ie. chargino, neutralino or slepton) in the R-parity conserved scenario gives the advantage of a clean signature, 
up to four leptons and
high missing tranverse energy. Fig.\ref{electrowikino_limit} summarises the different limits obtained in CMS, depending on the decay
chain via a slepton or a W/Z boson \cite{cms_ewkino}.

\indent
R-parity can be also violated (RPV). If only one among the leptonic and baryonic violating terms in the lagrangian is non-zero, then there is no 
problem with the proton decay. In this case, the missing transverse energy due to the LSP is suppressed. The final state will encompass several
jets, including b-jets and leptons. In this RPV configuration, the stop mass is probed up to 1.0 TeV, the gluino mass up to 1.5 TeV and the
squark mass up to 1.7 TeV using the 8 TeV data in ATLAS \cite{atlas_rpv} and CMS \cite{cms_rpv}.

\indent
The prospects for SUSY searches have been performed with both assumptions of 300 /fb (presumably obtained around 2022) and 3000 /fb (high luminosity LHC 
program, around 2030). The center-of-mass energy in this case is 14 TeV. The SUSY cross sections will benefit from this improvement as shown 
on Fig.\ref{xsec_prospects}. Fig.\ref{gluino_prospects} shows the ATLAS expected exclusion limit in the gluino-squark masses plane, as well as the discovery reach for
both targeted integrated luminosities. The expected sensitivity area will be largely 
extended \cite{atlas_susy_prospects}.

\begin{2figures}{hbtp}
  \resizebox{\linewidth}{0.8\linewidth}{\includegraphics{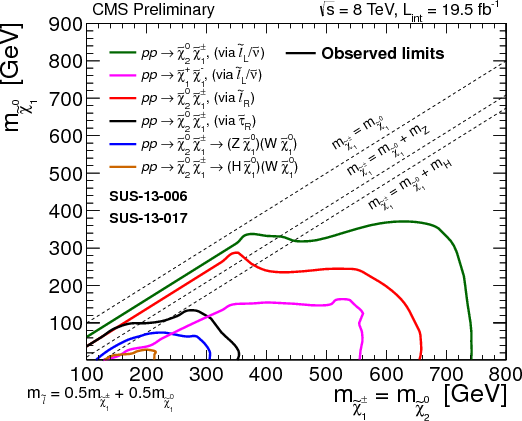}} &
  \resizebox{\linewidth}{0.8\linewidth}{\includegraphics{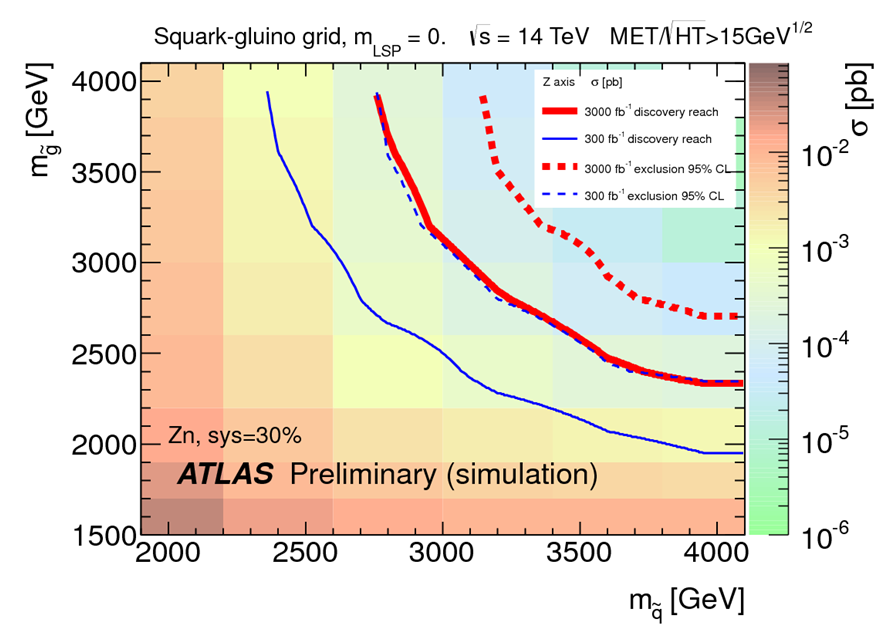}} \\
  \caption{Electroweakino limits obtained in CMS using all 7 TeV and 8 TeV data \cite{cms_ewkino}.}
  \label{electrowikino_limit} &
  \caption{Expected gluino mass limit using 300 /fb and 3000 /fb at 14 TeV in ATLAS \cite{atlas_susy_prospects}.}
  \label{gluino_prospects} \\
\end{2figures}

\section{Other searches beyond the SM}

\indent
The other searches for physics beyond the SM, also called Exotica program, are covering a wide range of analyses looking for additional 
heavy gauge bosons (W',Z'), extra-dimensions, compositeness, dark matter, also microscopic black holes, or any other 
scenario outside the SM (hidden valleys, unparticles ...). The typical topologies in Exotica imply some high tranverse momentum objects 
(electron, muon) at the TeV scale, high mass dilepton, diphoton or dijet resonances, or multi-lepton anomalous production.

\subsection{High transverse momentum leptons}

\indent
Several analyses in Exotica are based on high transverse momentum lepton searches. For instance the heavy gauge boson W' mass is probed up
to 3.3 TeV in CMS in the sequential SM, assuming the same branching ratio as in the SM. Fig.\ref{cms_wprime} shows the results combining 
both electron and muon channels \cite{cms_wprime}. 

\begin{2figures}{hbtp}
  \resizebox{\linewidth}{0.9\linewidth}{\includegraphics{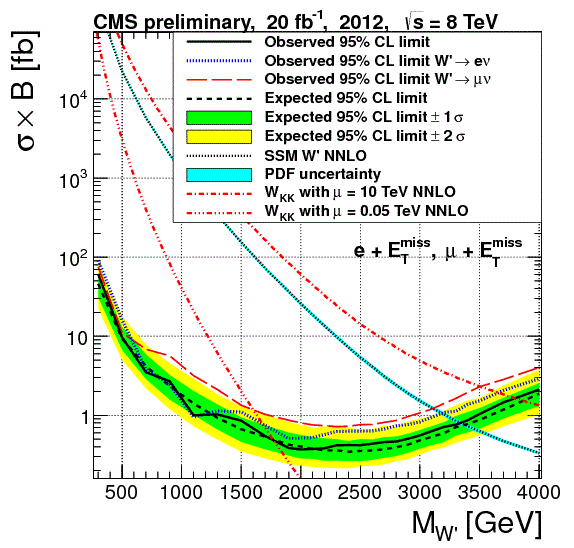}} &
  \resizebox{\linewidth}{0.9\linewidth}{\includegraphics{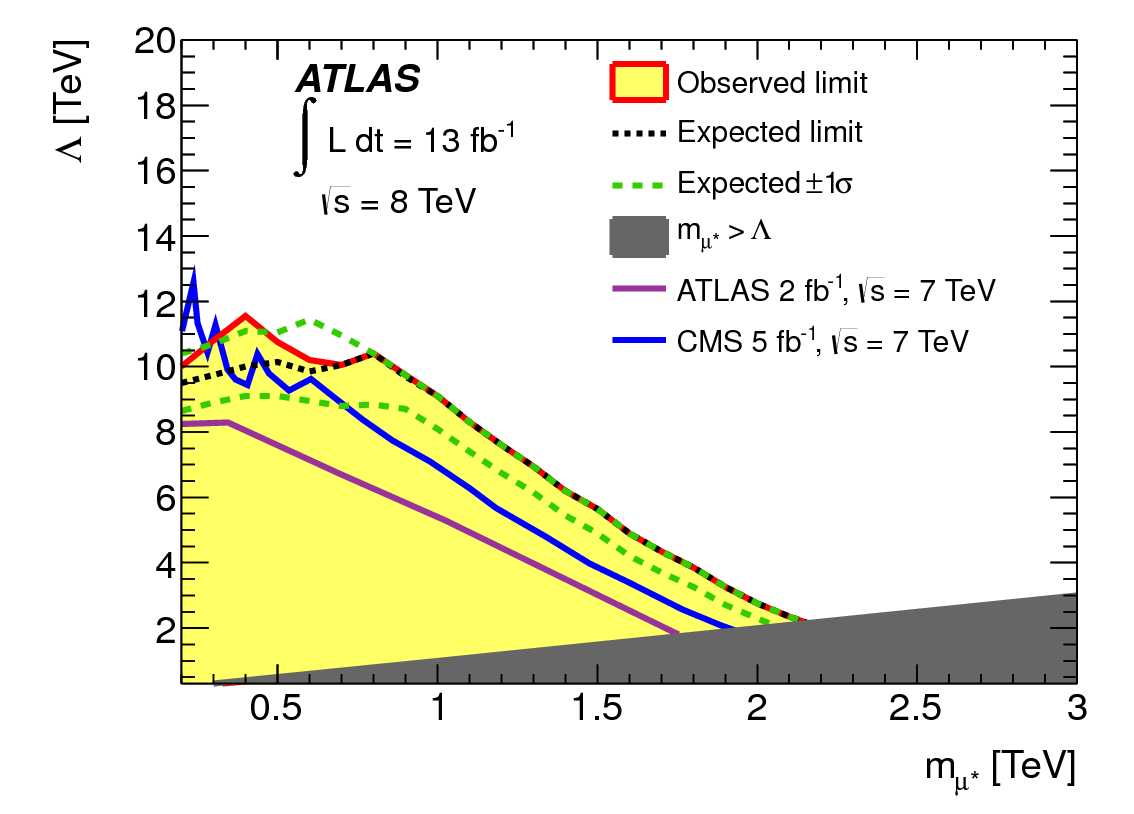}} \\
  \caption{W' searches in CMS in the electron and muon channels using the 8 TeV data \cite{cms_wprime}.}
  \label{cms_wprime} &
  \caption{Excited lepton searches in ATLAS using the 8 TeV data \cite{atlas_mu_star}.}
  \label{atlas_mu_star} \\
\end{2figures}

\indent
The searches for excited electrons or muons are also using high transverse momentum leptons. These excited states could be the
consequence of fermion substructure (compositeness). The typical decays in this analysis would be a lepton and a gauge boson or a lepton
and a pair of fermions. The probe of the mass scale of excited leptons is up to 2.2 TeV in ATLAS \cite{atlas_mu_star}. 
Fig.\ref{atlas_mu_star} shows the searches in ATLAS using the 8 TeV data and the comparison with previous results at 7 TeV in 
ATLAS and CMS.

\subsection{High mass resonances}

\indent
Fig.\ref{high_mass_ee} and Fig.\ref{high_mass_jets} show two high mass resonance distributions, respectively the di-electron and
the dijet spectrum. The sum of the SM backgrounds and the data are in good agreement, as shown also on the ratio plot below each distribution. 
Then a limit can be derived for several Exotica models. 

\begin{2figures}{hbtp}
  \resizebox{\linewidth}{0.9\linewidth}{\includegraphics{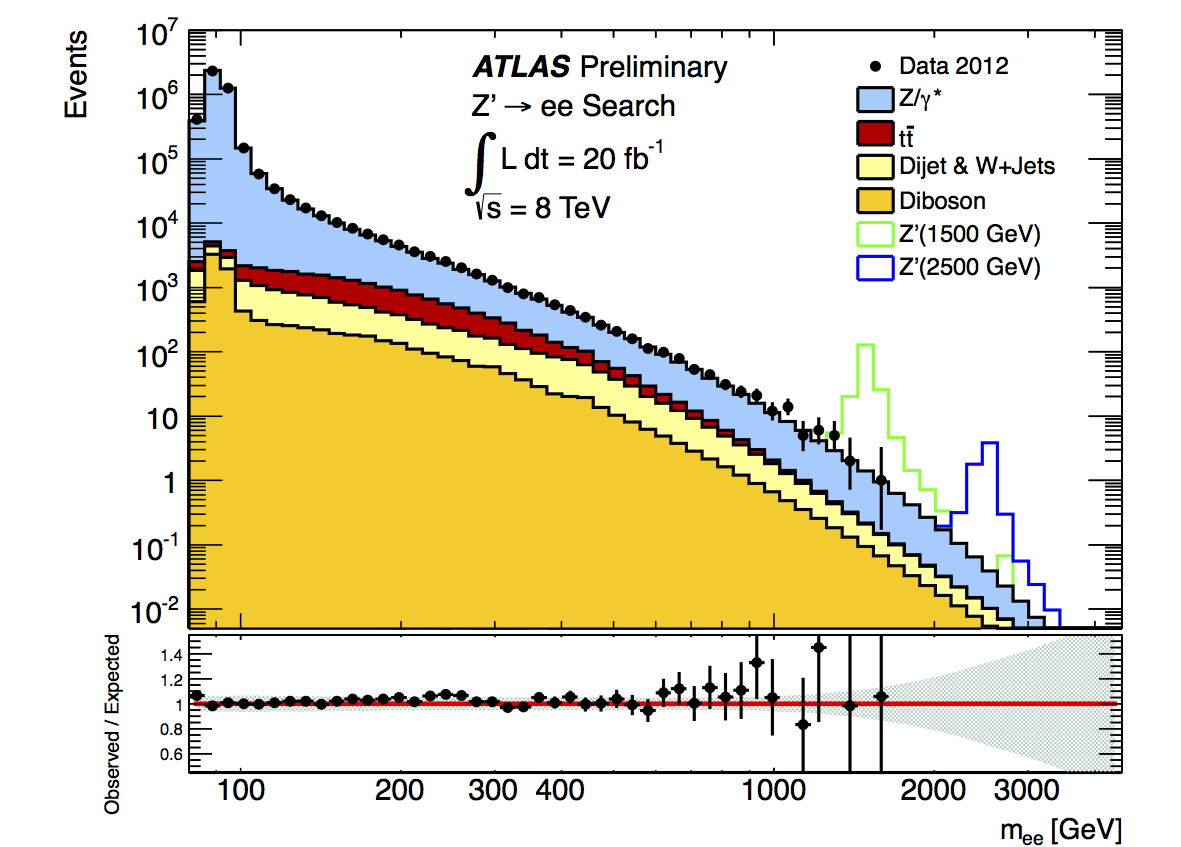}} &
  \resizebox{\linewidth}{0.9\linewidth}{\includegraphics{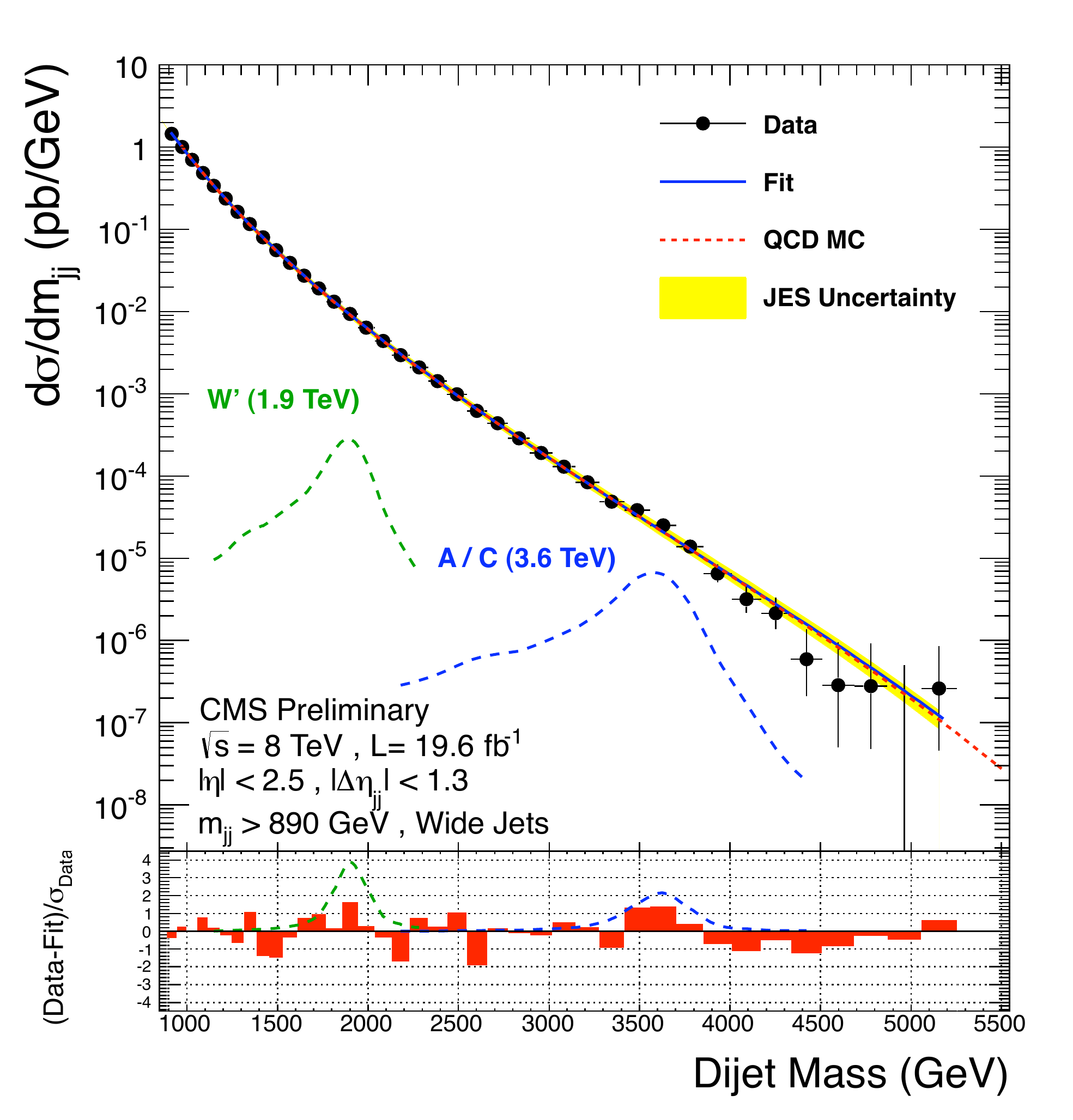}} \\
  \caption{Di-electron mass spectrum in ATLAS using the 8 TeV data \cite{atlas_zprime}.}
  \label{high_mass_ee} &
  \caption{Dijet mass spectrum in CMS using the 8 TeV data \cite{cms_dijet}.}
  \label{high_mass_jets} \\
\end{2figures}

\indent
The di-lepton (electron or muon) resonances allow first of all to put a limit on the Z' boson mass. The sequential SM 
(same coupling to fermions as the Z boson) being used 
as benchmark, the lower mass limit obtained in ATLAS is 2.9 TeV \cite{atlas_zprime}.  The extra-dimensions are probed by 
using the di-lepton and di-photon final states as shown on Fig.\ref{cms_dielec}. Depending on the number of extra-dimensions, the
mass scale has been probed up to 3 TeV for most of the scenarios \cite{cms_extra_dielec}. Fig.\ref{cms_dijet} shows the different limits
obtained from the dijet mass spectrum interpretation in the following models: strings, excited quarks, axigluon/colorons, W', Z' and also 
Randall-Sundrum gravitons. The lower mass limit reached in CMS is up to 5.1 TeV \cite{cms_dijet}.

\begin{2figures}{hbtp}
  \resizebox{\linewidth}{0.9\linewidth}{\includegraphics{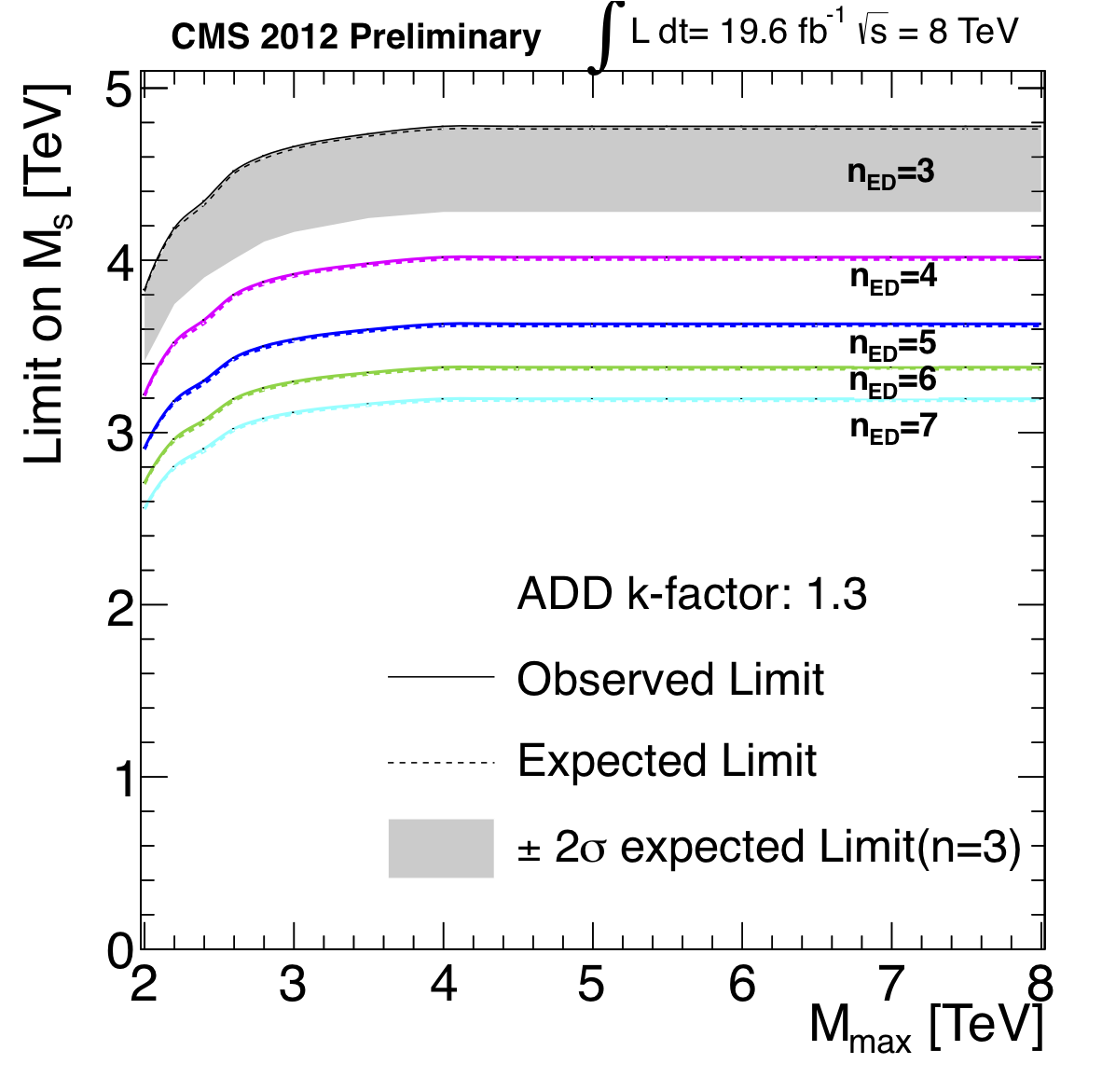}} &
  \resizebox{\linewidth}{0.9\linewidth}{\includegraphics{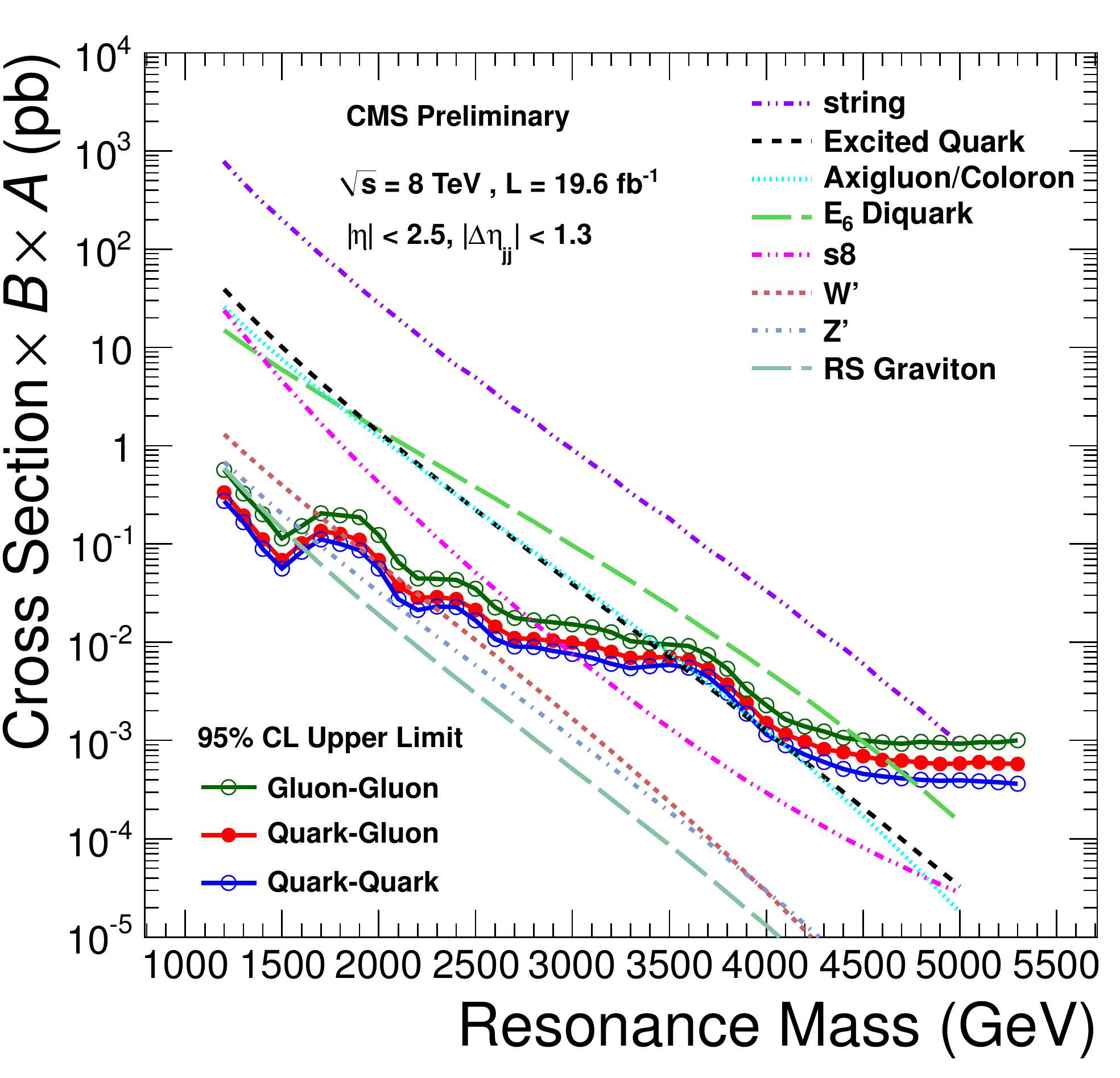}} \\
  \caption{Lower limit from CMS on the model parameter M$_{S}$ in an extra spatial dimension model using the 8 TeV data, depending on 
  the number n$_{ED}$ of extra dimensions \cite{cms_extra_dielec}.}
  \label{cms_dielec} &
  \caption{Search for narrow resonances using the dijet mass spectrum at 8 TeV in CMS \cite{cms_dijet}.}
  \label{cms_dijet} \\
\end{2figures}

\subsection{Anomalous lepton and top production}

\indent
Several inclusive model independent analyses are based on searches for anomalous multi lepton production. Excited neutrino,
fourth-generation quark or doubly-charged Higgs boson models are predicting anomalous multiple (at least three) lepton 
production. The scalar sum of the transverse momentum of the three charged leading leptons is compared to the SM expectation and 
an upper limit on the event yields due to non-SM processes is derived \cite{atlas_multi_lept}. 
Fig.\ref{atlas_multi_leptons} is summarizing these searches in ATLAS. 

\indent
The Kaluza-Klein (KK) excitations of gluons or gravitons in the extra-dimensions models, as well as new heavy gauge boson could have 
enhanced couplings to top anti-top pairs. Generic searches for anomalous production in the top/anti-top invariant mass spectrum 
allowed to put some new lower limits on the Z' mass (2.1 TeV) and also on the Randall-Sundrum KK gluons (2.5 TeV) using the 8 TeV data in CMS \cite{cms_b2g}.
Fig.\ref{cms_tt} shows the upper limit on the Z' cross section obtained in CMS, taking into account the branching ratio of the Z' into
top/anti-top pairs.

\begin{2figures}{hbtp}
  \resizebox{\linewidth}{0.8\linewidth}{\includegraphics{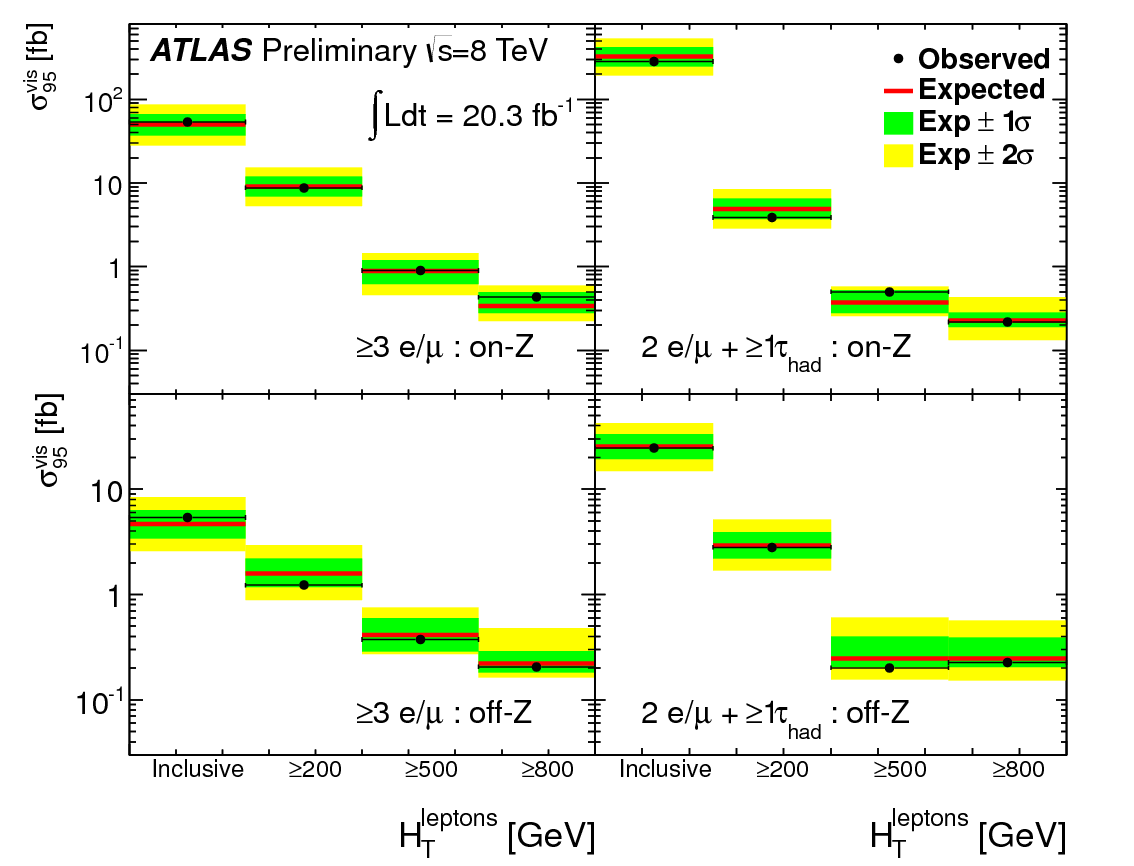}} &
  \resizebox{\linewidth}{0.8\linewidth}{\includegraphics{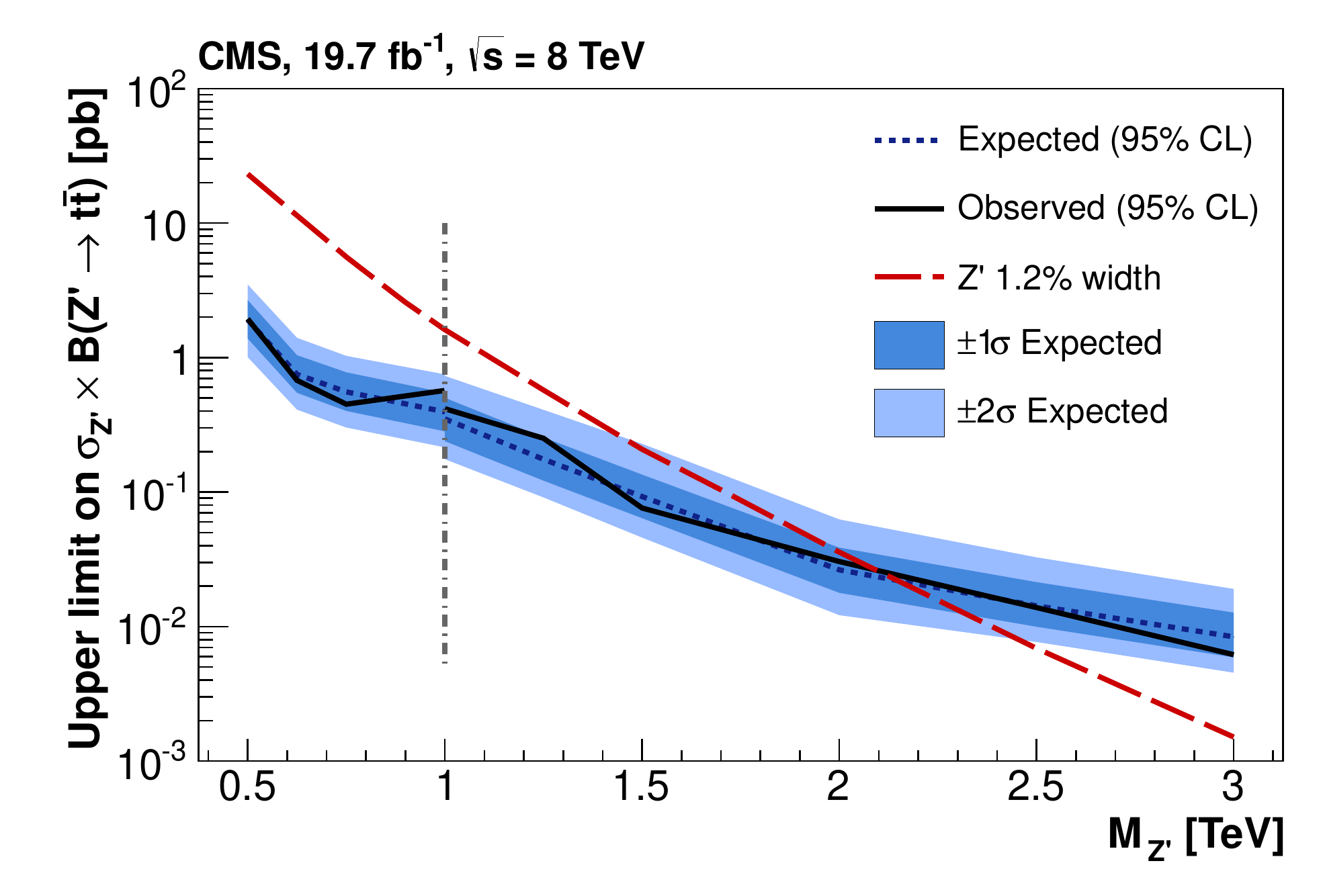}} \\
  \caption{Limit on the visible cross sections for several signal channels as a function of the lower bounds on the scalar sum of 
  the transverse momentum of the leptons  \cite{atlas_multi_lept}.}
  \label{atlas_multi_leptons} &
  \caption{Limits on the production cross section times branching fraction for a Z' using the top anti-top mass spectrum at 8 TeV in CMS \cite{cms_b2g}.}
  \label{cms_tt} \\
\end{2figures}

\subsection{Dark matter searches}

\indent
The dark matter-nucleon scattering cross section can be constrained in both spin dependent or independent interaction models 
\cite{atlas_dark_matter}. The collider experiments are sensitive in particular to the low mass region below 10 GeV.
Fig.\ref{atlas_dark_matter1} and Fig.\ref{atlas_dark_matter2} show the ATLAS searches for dark matter, as well as the comparison 
with the direct detection experiments (XENON100, CoGeNT, CDMS, COUP ...). This analysis is based on the assumption of Weakly Interacting Massive Particle (WIMP) pair
production via an unknown intermediate state and an initial radiation of a W or Z boson. The results can be also interpreted to set an
upper limit on the cross section of the Higgs boson decaying into invisible particles.

\begin{2figures}{hbtp}
  \resizebox{\linewidth}{0.8\linewidth}{\includegraphics{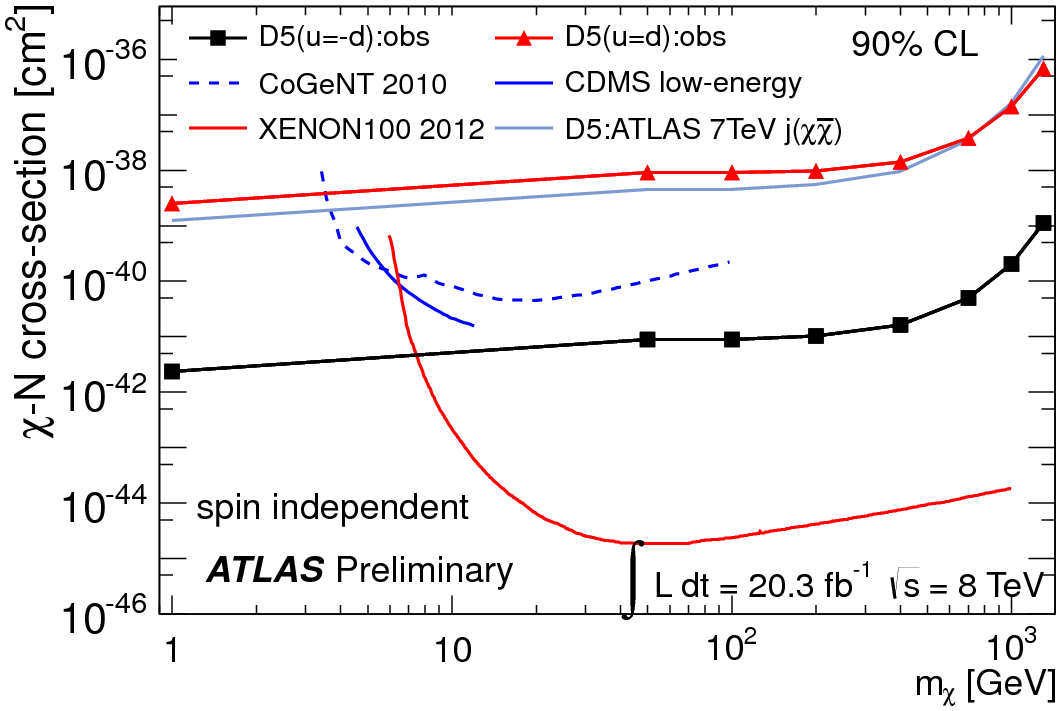}} &
  \resizebox{\linewidth}{0.8\linewidth}{\includegraphics{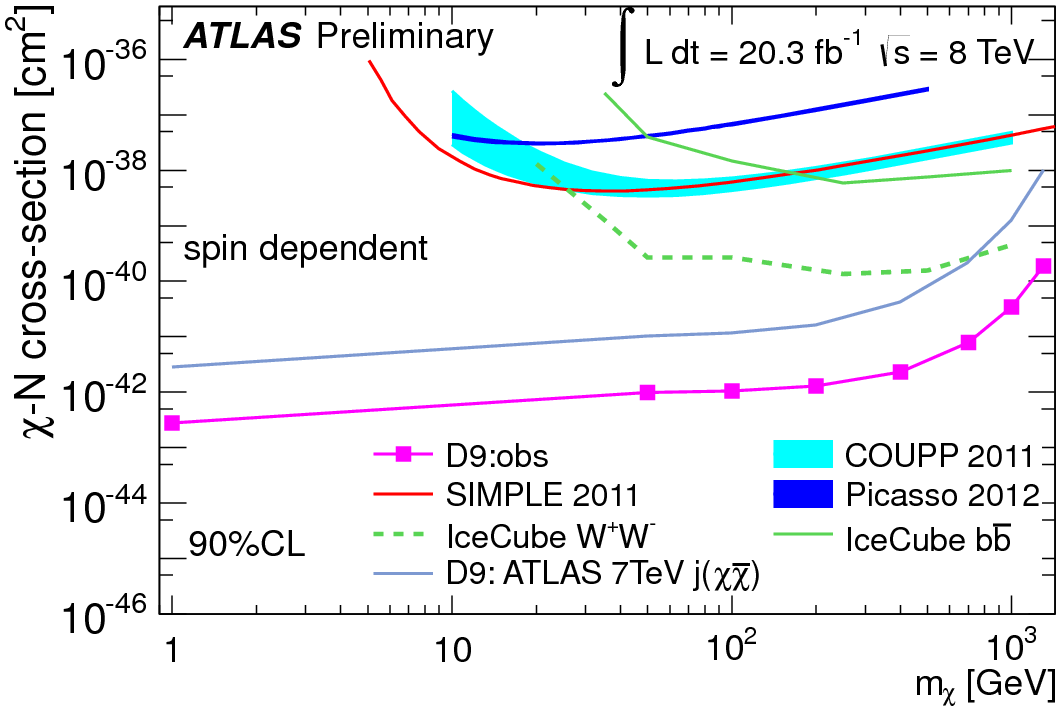}} \\
  \caption{Dark matter searches in spin independent model in ATLAS using 8 TeV data \cite{atlas_dark_matter}.}
  \label{atlas_dark_matter1} &
  \caption{Dark matter searches in spin dependent model in ATLAS using 8 TeV data \cite{atlas_dark_matter}.}
  \label{atlas_dark_matter2} \\
\end{2figures}

\section{Conclusion}

\indent
So far no hints of new physics beyond the SM have been found at the LHC. The SUSY searches in both ATLAS and CMS allow to probe the gluino mass up
to around 1.3 TeV while the squark masses are probed up to roughly 700 GeV. SUSY models with R-parity violated have been also investigated.
The exotica program has not shown any excess and limits are set on the mass scale for a lot of scenarios. The inclusive model
independent analyses permit to widen the range of searches, without any bias on the expected signal. Both SUSY and Exotica searches will
benefit from the higher center-of-mass energy and the increased luminosity in the next years and afterwards, using the HL-LHC program.

\section*{Acknowledgments}
This work was supported by the BMBF (Bundesministerium f\"{u}r Bildung und Forschung).

\end{document}